\documentclass[aps,preprint]{revtex4}%
\usepackage{amsfonts}
\usepackage{amsmath}
\usepackage{amssymb}
\usepackage{graphicx}%
\begin{document}
\title{\textbf{INFLATIONARY PHASE IN A GENERALIZED BRANS-DICKE THEORY}\\International Journal of Theoretical Physics, to be published.\\DOI 10.1007/s10773-009-9965-5}
\author{Marcelo S.Berman$^{(1)}$}
\author{and Luis A.Trevisan$^{(2)}$}
\affiliation{$^{1}$Instituto Albert Einstein / Latinamerica - }
\affiliation{Av. Candido Hartmann, 575 - \ \# 17}
\affiliation{80730-440 - Curitiba - PR - Brazil }
\affiliation{email: msberman@institutoalberteinstein.org\\}
\affiliation{$^{2}$Universidade Estadual de Ponta Grossa,}
\affiliation{Demat, CEP 84010-330, Ponta Grossa,Pr,}
\affiliation{Brazil \ email: luisaugustotrevisan@yahoo.com.br}

\begin{abstract}
We find a solution for exponential inflation in a Brans-Dicke \ generalized
\ model, where the coupling \textquotedblleft constant \textquotedblright\ is
\ variable. While \ in General \ Relativity the equation of state is
\ \ \ $p=-\rho,$ \ \ here we find $\ \ \ p=\alpha\rho,$ \ where $\alpha<-2/3.
$ \ The negativity of cosmic pressure implies acceleration of the expansion,
even with \ \ $\Lambda<0$\ \ \ .


\end{abstract}
\maketitle

\newpage

\begin{center}
\bigskip\textbf{INFLATIONARY PHASE IN A }

\textbf{GENERALIZED BRANS-DICKE THEORY}

International Journal of Theoretical Physics, to be published.

DOI 10.1007/s10773-009-9965-5

MARCELO S. BERMAN and LUIS A. TREVISAN
\end{center}

{\large \bigskip Introduction}

New evidence for primordial inflation has been recently gathered through
cosmic microwave observation (Weinberg, 2008). Barrow (1993) has pointed out
the possible relevance of scalar-tensor gravity theories in the study of the
inflationary phase during the early Universe. He obtained exact solutions for
homogeneous and isotropic cosmologies in vacuum and radiation cases, for a
variable \ coupling \textquotedblleft constant \textquotedblright%
,$\omega=\omega(\phi),$ where $\phi$ stands for the scalar field. For accounts
on inflation, see, for instance, Linde's \ book (Linde, 1990).

In this letter we extend \ Barrow's paper by the study of an inflationary
exponential phase. This letter can be considered also as a complement \ to
Berman and Som's paper (Berman and Som, 1989) dealing with the inflationary
phase in B.D. original framework, which was followed by a letter by Berman
(1989) where he studied the same problem in the context of a B.D. theory
endowed with a cosmological constant. For scalar-tensor theories, consult the
books by Berman\ (2007), Faraoni (2004), and Fujii and Maeda (2003). In Berman
(2007a), we find a \textit{rationale} for the existence of a cosmological
"constant", though we must remember that a negative cosmic pressure may be
also responsabilized for accelerated expansion, which includes exponential inflation.

{\large \bigskip The Field Equations}

One way to formulate a scalar-tensor theory of gravitation can be with the
following Lagrangian:%

\begin{equation}
L_{\phi}=-\phi R+\phi^{-1}\omega(\phi)\partial_{a}\phi\partial^{a}\phi+16\pi
L_{m}-2\Lambda(\phi)
\end{equation}
where $L_{m}$ is the Lagrangian for matter fields, and $\phi$ is the scalar
field. If $\omega=const$ we obtain the Brans-Dicke (1961) \ theory. This
Lagrangian was adopted by Barrow and Maeda (1990). For a discussion about the
Lagrangians of the scalar theories of gravitation, see (Liddle and Wands,
1992). The cosmological term \ $\Lambda(\phi)$\ \ is taken also to mean
time-dependent lambda.

By varying the action associated with (1) with respect to the space-time
metric and the scalar field $\phi,$ respectively we obtain the generalized
Einstein equations and the wave equation for $\phi$ (Barrow, 1993):%

\begin{equation}
G_{ab}=-\frac{8\pi}{\phi}T_{ab}-\frac{\omega}{\phi^{2}}\left[  \phi_{a}%
\phi_{b}-\frac{1}{2}g_{ab}\phi_{i}\phi^{i}\right]  -\frac{1}{\phi}\left[
\phi_{a;b}-g_{ab}\Box\phi\right]  -\frac{\Lambda}{\phi}g_{ab}%
\end{equation}

\begin{equation}
\left[  3+2\omega\right]  \Box\phi=8\pi T-\left(  \frac{d\omega}{d\phi
}\right)  \phi_{i}\phi^{i}+2\phi\frac{d\Lambda}{d\phi}-4\Lambda
\end{equation}
\ \ \ 

\bigskip

In General Relativity theory, in face of a perfect fluid matter field, from
the field equations, it is derived the energy momentum conservation law,%

\begin{equation}
T_{;b}^{ab}=0.
\end{equation}

\bigskip In Brans-Dicke theory, Weinberg (1972) has commented that in order to
preserve the Principle of Equivalence, the scalar-field does not enter into
the conservation equation above, which takes into consideration only the
matter-fields. For scalar-tensor theories, as well, this conservation equation
is imposed on the same token, but, of course, if we take the field equations,
say, for Robertson-Walker's metric, obtaining an equation for cosmic pressure
and other for the energy density, we could combine those equations, along with
the scalar-field one, and obtain a generalisation of the kind,

$\ \ \ \ \ \ \ \ \ \ \ \ \ \ \ \ \ \ \ \ \ \ \ \ \ \ \ \ \ \ \ \ \ \ \ \ \ \ \ \ \ \ \ \ \ \ \ \ G_{;b}%
^{ab}=0$ \ \ \ \ \ \ \ \ \ \ \ \ \ \ \ \ \ \ \ ,

\bigskip

where the conservation law applies to the right-hand-side of \ (2).

With Robertson-Walker's metric,%

\begin{equation}
ds^{2}=dt^{2}-a^{2}\left[  \left(  1-kr^{2}\right)  ^{-1}dr^{2}+r^{2}%
d\theta+r^{2}\sin^{2}\theta d\varphi^{2}\right]
\end{equation}
we find, from (4), (2) and (3), that :%

\begin{equation}
\frac{\dot{a}^{2}}{a^{2}}+\frac{k}{a^{2}}=\frac{8\pi\rho}{3\phi}-\frac
{\dot{\phi}}{\phi}\frac{\dot{a}}{a}+\frac{\omega}{6}\frac{\dot{\phi}^{2}}%
{\phi^{2}}+\frac{\Lambda}{3\phi}%
\end{equation}

\begin{equation}
\dot{\rho}+3\frac{\dot{a}}{a}(\rho+p)=0
\end{equation}

\begin{equation}
\ddot{\phi}+\left[  3\frac{\dot{a}}{a}+\frac{\dot{\omega}}{2\omega+3}\right]
\dot{\phi}=\frac{1}{3+2\omega}\left[  8\pi(\rho-3p)-2\phi\frac{\dot{\Lambda}%
}{\dot{\phi}}+4\Lambda\right]  \text{ \ },
\end{equation}
where overdots stand for time derivatives.

From now on, we consider only spatially flat solutions ($k=0$). Let%

\begin{equation}
a=a_{0}e^{Ht}%
\end{equation}
where $a_{0,}H,$ are constants, and%

\begin{equation}
p=\alpha\rho
\end{equation}
($\alpha=const).$

In General \ Relativity, $\alpha=-1$ for the inflationary phase; here, we must
consider also other possibilities. Law (10) stands for a \textquotedblleft
perfect \textquotedblright\ gas equation of state. From (7) and (9), we find,
using (10),%

\begin{equation}
\rho=\rho_{0}e^{-3H(1+\alpha)t}%
\end{equation}
where $\rho_{0}=const.$

Remember that $H=\dot{a}/a$ stands for Hubble's parameter.

Consider the solution for $\phi(t)$, and $\omega(t)$:%

\begin{equation}
\frac{\dot{a}^{2}}{a^{2}}=-\frac{\dot{\phi}}{\phi}\frac{\dot{a}}{a}\text{
\ \ \ \ \ \ \ \ \ \ \ \ \ \ \ \ \ \ \ \ ,}%
\end{equation}

\bigskip and,%

\begin{equation}
\frac{8\pi\rho}{\phi}=-\frac{\omega}{6}\frac{\dot{\phi}^{2}}{\phi^{2}}%
-\frac{\Lambda}{3\phi}%
\end{equation}

\bigskip

\bigskip By summing (12) and (13), we recover expression (6), so that the
above two equations, result in a particular solution, though that it has some
generality altogether.

We find from (12),%

\begin{equation}
\phi(t)=\phi_{0}e^{-Ht}%
\end{equation}

\bigskip

and,

\bigskip

$\qquad\qquad\qquad\qquad\qquad\qquad\qquad\Lambda=\Lambda_{0}e^{-3H(1+\alpha
)t}$ \ \ \ \ \ \ \ \ \ \ ,

\bigskip

with \ $\phi_{0}$ , $\ \Lambda_{0}\ \ =$ constants $.$

\bigskip

From (13), we get a possible solution with, \ $\omega\gg3/2,$ \ %

\begin{equation}
\omega\cong\omega_{0}e^{-H(2+3\alpha)t}%
\end{equation}
with \ $\omega_{0}=$ \ positive constant $\ $, \ and $\phi_{0}>0.$

\bigskip

The constants must obey the condition, obtained from (12) and (13),

\bigskip

$\ \ \ \ \ \ \ \ \ \ \ \ \ \ \ \ \ \ \ \ \ \ \ \ \ \ \ \ \ \ 8\pi\rho
_{0}+\frac{1}{3}\Lambda_{0}+\frac{1}{6}H^{2}\phi_{0}\omega_{0}=0$ \ \ \ \ \ \ \ \ \ \ \ \ \ \ \ .

\bigskip

\bigskip The reason for a positive scalar-field, is that gravitation should be
kept attractive, i.e., Newton's gravitational constant is positive. The reason
for a positive coupling "constant", is that experimental gravitational and
astronomical observations require a large positive \ $\omega$\ \ value.

\bigskip

It is then, highly desirable that $\omega$ grow with time, so we impose,%

\begin{equation}
\alpha<-\frac{2}{3}%
\end{equation}
\ This condition on the equation of state encompasses the case $\alpha=-1$ of G.R.

\bigskip

From the scalar-field equation, of course, we get the fulfilled approximate
condition, \bigskip\bigskip%

\begin{equation}
2H^{2}\phi_{0}\cong\frac{M}{\left(  3+2\omega\right)  }e^{-H\left(
2+3\alpha\right)  t}\text{ \ \ \ \ \ \ \ \ \ \ \ \ \ \ \ \ \ ,}%
\end{equation}

\bigskip

\bigskip where, \ \ 

\ %

\begin{equation}
M=\omega_{0}H^{2}\phi_{0}^{2}\left(  2+3\alpha\right)  +4\Lambda_{0}+6\left(
1+\alpha\right)  \Lambda_{0}-8\pi\rho_{0}\left(  1-3\alpha\right)  \
\end{equation}

\bigskip

{\large \bigskip Conclusion}

We have thus, found a new solution for inflation, that deserves attention. On
the other hand, it can be shown (Berman and Trevisan, 2009) that condition
(16) is necessary for the amplification of gravitational waves during
exponential inflation, at least when \ $\omega$ is constant.

\bigskip

It could be argued that another possible solution would be given by a positive
cosmological "constant", followed by a negative coupling \ $\omega$\ \ . \ It
certainly may be more palatable for string theorists, but it would require
some hand waving of the type \ $\omega_{0}<-2$\ \ , because otherwise, the
observation of slowly moving particles or of time-dilation experiments,
\ which imply (see Weinberg, 1972),

\bigskip%

\begin{equation}
G=\left[  \frac{2\omega+4}{2\omega+3}\right]  \phi^{-1}\text{
\ \ \ \ \ \ \ \ \ \ \ \ ,}%
\end{equation}

\medskip would again carry \ \ $G<0$\ \ \ .

\bigskip

\bigskip\bigskip{\Large Acknowledgments}

\bigskip MSB thanks two anonymous referees of this journal, who contributed
significantly towards the preparation of the final manuscript. MSB is also
grateful to Nelson Suga, Marcelo F. Guimar\~{a}es, Antonio F. da F. Teixeira,
and Mauro Tonasse and for the encouragement by Albert, Paula, and Geni. He
offers this paper \textit{in memoriam of \ \ }M. M. Som.

{\Large References}

\bigskip

Barrow, J.D;--Phys. Rev D \textbf{46}, 5329 (1993)

Linde, A-``Particle Physics and Inflationary Cosmology '', Harwood Acad.
Press, N.Y. (1990).

Berman, M.S; Som, M.M; -- Phys.Lett.\textbf{\ }A\textbf{\ 136}, 206, (1989)

Berman,M.S -- Phys.Lett. A\textbf{\ 142}, 335,(1989)

Brans, C. , Dicke, R.H;-- Phys.Rev. \textbf{124,} 925 (1961)

Barrow,J.D. and Maeda, K.--Nucl.Phys \textbf{B341,} 294 (1990)

Liddle, A.R. and Wands, D--Phys.Rev. \textbf{D45,}2665 (1992)

Berman,M.S, Trevisan, L.A; -- Submitted.

Weinberg, S. -- \textit{Cosmology, }Oxford University Press, Oxford (2008)

Berman, M.S. -- \textit{Introduction to General Relativistic and Scalar-Tensor
Cosmologies, }Nova Science Publishers, New York (2007).

Faraoni, V. -- \textit{Cosmology in Scalar-Tensor Gravity, }Kluwer, Dordrecht (2004).

Fujii, Y.; Maeda, K. -I. -- \textit{The Scalar-Tensor Theory of Gravitation,
}CUP, Cambridge (2003).

Berman, M.S. - \textit{Introduction to General Relativity and the Cosmological
Constant Problem, }Nova Science Publishers, New York (2007a).

Weinberg, S. - \textit{Gravitation and Cosmology, }Wiley, New York (1972).
\end{document}